\documentstyle[epsfig]{elsart}

\begin{document}

\begin{frontmatter}

\title{The RHIC Zero Degree Calorimeters} 

\author[IKF]{C. Adler},
\author[IHEP]{ A. Denisov}, 
\author[UMD]{E. Garcia}, 
\author[TAM]{M. Murray}, 
\author[IKF]{H. Stroebele}, 
\author[BNL]{S. White}

\address[BNL]{\it Brookhaven National Lab, Upton, NY  11973 USA}  
\address[IHEP]{\it IHEP-Protvino,Russia}                          
\address[IKF]{\it IKF ,University of Frankfurt,Germany}                                
\address[TAM]{\it Texas A\&M Cyclotron,College Station, TX 77843} 
\address[UMD]{\it U. of Maryland, College Park,Md, USA}           

\end{frontmatter}

\section{Introduction}

High Energy collisions of nuclei usually lead to the emission of evaporation
neutrons from both ``beam'' and ``target'' nuclei. At the RHIC heavy
ion collider with 100GeV/u beam energy, 
evaporation neutrons  diverge by less than $~2$ milliradians from the beam axis.
Neutral beam fragments can be detected downstream of RHIC ion
collisions (and a large aperture Accelerator dipole magnet) 
if $\theta\leq$ 4 mr\  but charged fragments in the same
angular range are usually too close to the beam trajectory.
	In this 'zero degree' region produced
particles and other secondaries deposit negligible  
energy when compared with that of beam fragmentation neutrons.

        The purpose of the RHIC zero degree calorimeters (ZDC's) is to detect
neutrons emitted within this cone along both beam directions and measure their
total energy (from which we calculate multiplicity). The ZDC coincidence of the 2 beam
directions is a minimal bias selection of heavy ion collisions. This makes
it useful as an event trigger and a luminosity monitor\cite{baltz} and 
for this reason we built identical detectors for all 4 RHIC experiments.
  The 
neutron multiplicity is also known to be correlated with event geometry
\cite{appel} and will be used to measure collision centrality in mutual beam interactions.

\section{Design Goals}

        The RHIC ZDC's are hadron calorimeters. Their longitudinal segmentation
(2 $\lambda_{I}$, $50 X_0$) is determined by practical, mechanical considerations.
Electro-magnetic energy emission into this region is predicted to be
negligible so this measurement is not emphasized in
our design.
 Since the spatial distribution of neutrons emitted in the fragmentation
region carries only limited information about the collision, the calorimeters
are built without transverse segmentation.

\begin{figure}
 \centerline{\psfig{figure=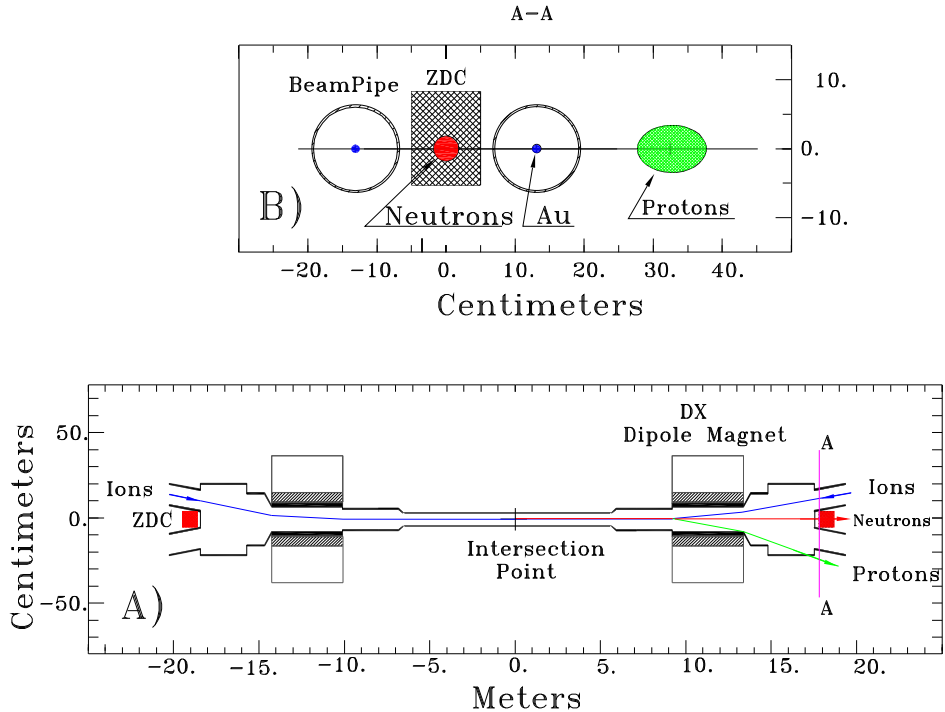,width=6in}}
  \caption[IR]{ { \label{fg:IR}}
       Plan view of the collision region and (section A-A)  "beam's eye" view of the
zdc location indicating deflection of protons and charged fragments
( with Z/A$\sim~1$ downstream of the "DX"
Dipole magnet. }
\end{figure}
        The Forward Energy resolution goal was determined by the
need to cleanly resolve the single neutron peak in peripheral
nuclear collisions. The         
natural energy spread of emitted single neutrons\cite{baltz} being 
approximately $ 10 \%$  a resolution of$\frac{\sigma_{E}}{E}\leq 20\%$
 at $E_{n}${}$=100$ GeV\null appeared reasonable.

The limited available space between the RHIC beams at the ZDC location
imposes the most stringent constraint on the calorimeter design. 
As can be seen from
Figure~1, the total width of the calorimeters is only cannot exceed 10 cm
(equal to 1
nuclear interaction length ($\Lambda_I$) in tungsten).
	We designed the ZDC's to minimize the loss in energy resolution due to
   shower leakage, which can cause fluctuation in measured shower
        energy through dependence on position of impact and random
        fluctuations in shower development. 
    
	Finally, the ZDC's are required to withstand a dose of 
$\sim10^5$ rad.,
which is the expected exposure during several years of RHIC 
operation\cite{AJS}.

\begin{figure}
 \centerline{\psfig{figure=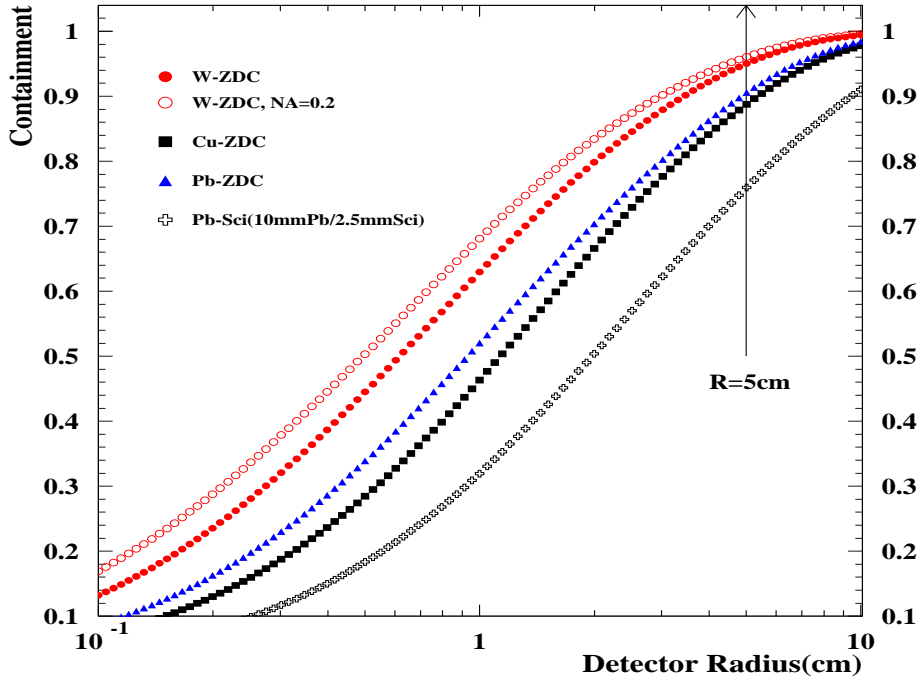,height=4in,width=5in}}
  \caption[con]{ { \label{fg:con}}
        Apparent shower profile for different hadron calorimeters. The W and Cu
parameters are described in the text. A variant with numerical aperture=0.2
is shown for comparison. }
\end{figure}

\section{Simulations}

	We simulated shower development,
        light production and transport in the optical components
        using Geant 3.21 \cite{geant} for 2 basic sampling calorimeter designs:
        
\begin{enumerate}

\item Pb absorber with scintillator sampling

\item Pb, Cu or $W$ absorber, each with undoped fiber optical ribbons in the
sampling layer 

\end{enumerate}

The ZDC sampling technique which we adopted for this project, is
sensitive to Cerenkov light produced by charged shower secondaries in a
commercial, PMMA based \cite{Toray} communication grade optical fiber
with characteristics given in a single free
parameter was used to match the (wavelength dependent) optical fiber attenuation 
coefficient and photomultiplier quantum efficiency to the observed 
signal from testbeam $\mu$ 's. Hadronic shower simulation is based on 
Geisha \cite{geant}
with a low energy cutoff of 0.5 MeV 
on electrons and photons and 1 MeV on hadrons.

\begin{table}
\caption{Mechanical parameters of the ZDC's}\label{Table 1}
\begin{tabular}{|l|l|l|l|}
\hline
       &     Absorber       & Space for fibers & Modules/Layers \\
\hline
Prototype W-ZDC  &      Tungsten      &      1.0 mm      &
4(8$\lambda_{I}$;218$X_0$)  \\
                 & (100x150x5 mm$^3$)  &                  &
27       \\
 \hline
Prototype Cu-ZDC &       Copper       &      1.0 mm      &
8(7.5$\lambda_{I}$;79$X_0$)  \\
                 & (100x150x10 mm$^3$) &                  &
10       \\
 \hline
 Production ZDC  &   Tungsten alloy   &      1.4 mm      &
3(5.1$\lambda_{I}$;149$X_0$) \\
                 & (100x187x5 mm$^3$)  &                  &
27       \\
 \hline
\end{tabular}
\end{table}
\begin{table}
\caption{Characteristics of the fiber ribbon material. NA${}=0.50$
 \label{tabone}} 
\begin{tabular}{|l|l|l|}
\hline
&      Outer Diam \qquad & Material/Index\\
\hline
 Core   &       0.45 mm     &
PMMA/1.49 \\ Cladding    &    0.50      &   fluorine doped/1.40 \\ Surface
Prep\qquad & 0.60  &       White Silicone Rubber/ EMA \\
\hline
\end{tabular}
\end{table}

	The fiber sampling layers, in all cases, consist of a single ribbon
 of 0.5mm diameter fibers  as shown in Fig.5. We
 chose an orientation of $45^o$ relative to the incident 
  beam direction which roughly coincides with the Cerenkov
 angle of $\beta$=1 particles in PMMA.
	PMMA fibers are readily available with a numerical aperture(NA) 
of 0.5 (defined as the fractional solid angle which is transmitted in
the fiber). This aperture corresponds to a maximum angle of $30^o$.
Quartz fibers generally have a smaller aperture.

In our simulations we studied:

\begin{enumerate}

\item the effects of transverse shower leakage

\item energy resolution dependence on sampling frequency and
photostatistics

\item dependence on fiber orientation

\end{enumerate}

Figure~2 illustrates the main advantage of the zdc(Cerenkov) vs.\ scintillator
sampling technique. We plot the fraction of the calorimeter signal
(photons transported in the scintillator or fiber) produced as a
function of radius from 100 GeV protons impacting the calorimeter. A radius
of 5 cm (the maximum space allowable at the RHIC location) contains 75\%
of the shower signal in the case of a Pb/Scintillator calorimeter with
10mm(Pb) and 2.5mm(Scint) layers. The same dimension Pb
calorimeter with Cerenkov sampling yields 91\% containment. In general the 
Cerenkov
technique with Pb absorber achieves a given level of containment with a 
factor of 2 smaller radius than with
scintillator.

Changing from Pb to $W$ absorber yields almost another a factor of  2
reduction in containment radius. On the other hand, reducing the fiber numerical aperture
in the sampling layer produces only a negligible change.


\subsection{Cerenkov light Production and Capture}

Because the optical fibers only transport Cerenkov light emitted nearly
aligned with the fiber axis, this detector is most sensitive to
charged particles which cross at approximately $45^o$ to the fiber axis.
 The lower energy shower component, which is more diffuse, is
therefore suppressed. 

	This filtering effect is reduced by multiple coulomb scattering 
of electrons and by the increased path length traversed by
particles with less than $45^o$ angle to the  fiber direction.

\begin{figure}
 \centerline{\psfig{figure=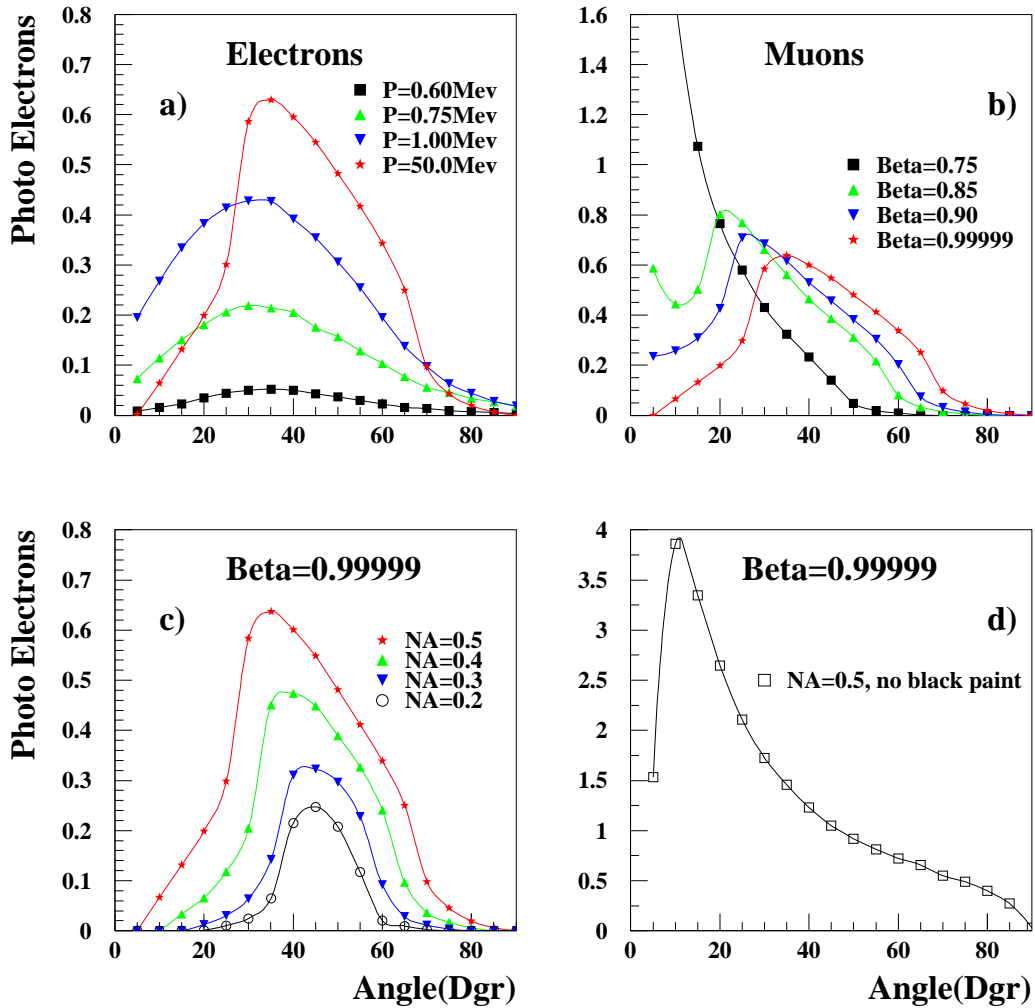,width=6in}}
  \caption[con]{ { \label{fg:con}}
        Photoelectron yield for 0.5 mm diameter PMMA fibers. }
\end{figure}

	In Figure~3 we plot the average photoelectron yields for muons and
electrons traversing a ribbon of 0.5mm diameter fibers as a
function of incidence angle. Comparing Figures~3a and 3b we see
 that multiple coulomb scattering has a significant effect on the response to low energy
 electrons. The improved angular filtering that could be
 obtained with lower NA fibers (see fig.~3c)
is largely offset by this effect and the lower light yield which
results from small aperture. 

Fig.~3d illustrates the effect of ``cladding modes'' which we
suppress by applying an ExtraMural Absorber (latex based black paint)
over the cladding layers of our fibers.

\subsection{Relative response to electrons and hadrons}

	Figure 4 shows the calculated response of our zdc modules to electrons
and protons as a function of beam energy. The response to beam
 energy (100 GeV) protons is a factor of 2 lower than for
 electrons of the same energy. In this sense, our design is
extremely non-compensating. This lack of compensation
is the dominant source of energy resolution of the calorimeter
for 100 GeV protons since the response changes with
fluctuations in the energy fraction carried by $\pi^o$'s in the
hadronic shower.	

\begin{figure}
 \centerline{\psfig{figure=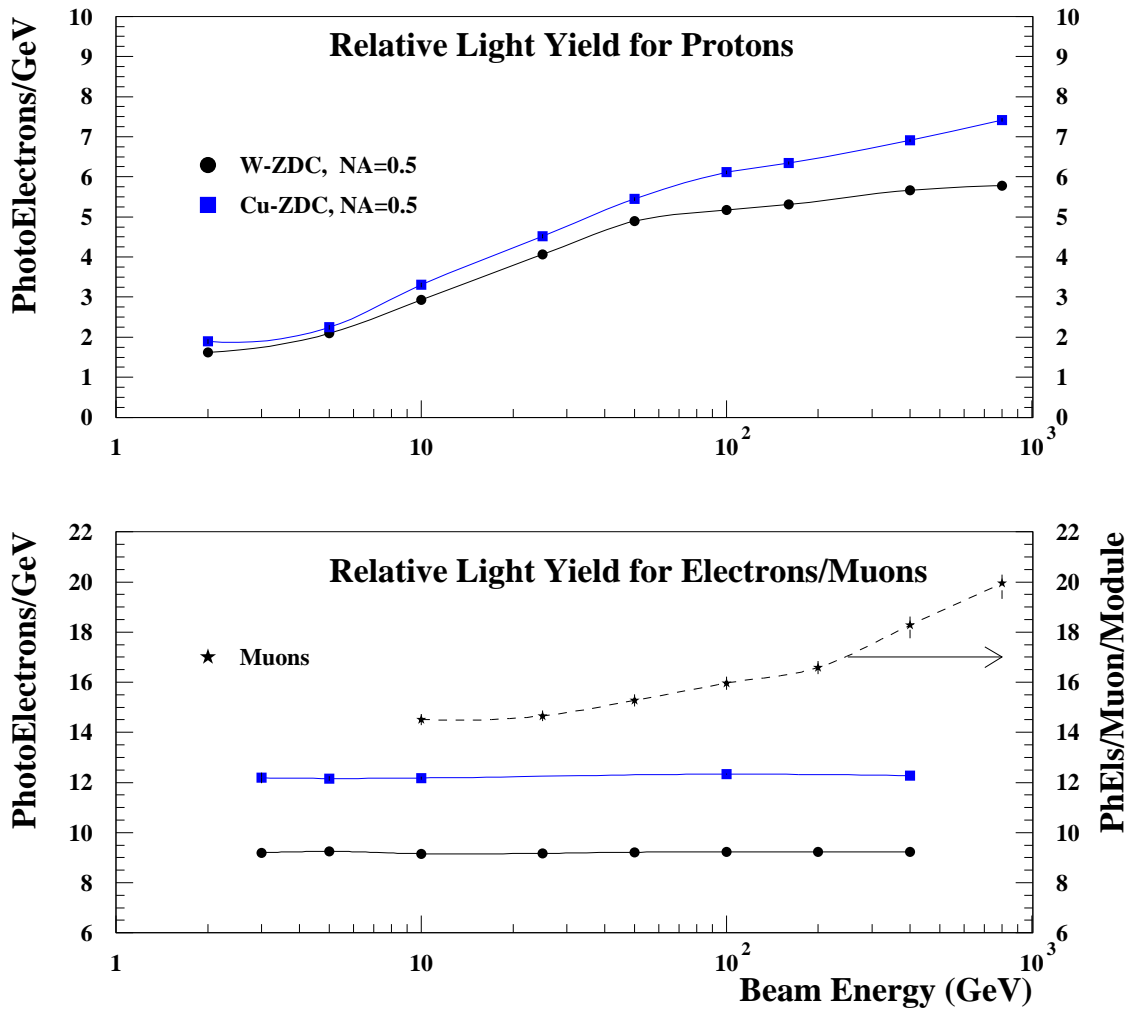,width=6in}}
  \caption[nonlinear]{ { \label{fg:nonlinear}}
        Calculated response of the ZDC to protons and electrons(for both 
Tungsten and Copper modules) and muons (shown for the tungsten modules).  }
\end{figure}

\subsection{Linearity}

	In our application, where the calorimeter is used to count beam energy
 neutrons, linearity is not a design consideration. It is clear
 from figure~4 that , while the response to electrons is linear
with energy, the hadron response is not. Response to neutrons
and protons approaches zero at low energies. Also shown in figure~4 is
the response to muons, which is energy independent up to $\sim100$ GeV/c at
which point radiative energy loss becomes significant. We use cosmic
ray and beam muons for detector pre-calibration. 

\subsection{Energy resolution}

The role of the main components of the energy resolution is illustrated
in Table~3, where we fit simulated calorimeter response to 50--800 GeV
proton induced showers to a stochastic plus a constant term. Our
results are a poor fit to a quadratic sum.

\begin{table}
  \caption{Calculated energy resolution of ZDC's}\label{Table 2}
\begin{tabular}{|l|l|l|l|l|}
 \hline
   Absorber& PhEls    & ``e/h" & Stochastic&Constant  \\
           &per 100Gev&ratio  & term(\%)   &term(\%)\\
 \hline
   W (2.5mm) &      1036    &  1.79        & 69.6$\pm$7.9         & 10.1$\pm$0.7\\
 \hline
   W(5.0mm)  &       518     & 1.78        & 84.6$\pm$4.8         & 9.1$\pm$0.5 \\
 \hline
   W(10mm)  &       256     & 1.78        & 92.4$\pm$8.2         & 8.8$\pm$0.6 \\
 \hline
   Cu(10mm) &       611     & 2.01        & 111.7$\pm$7.0        & 9.3$\pm$0.6 \\
 \hline
   Pb(10mm) &       422     & 1.80        & 91.0$\pm$8.9         & 9.5$\pm$0.6  \\
 \hline
\end{tabular}
\end{table}

If we were to increase the sampling frequency from 1/2 to twice our
design value of 1 per 5mm tungsten absorber there would be a negligible
change in resolution at 100 GeV\null. This configuration would reduce the 
stochastic
term due to photostatistics from 6 to 3\% but leave the dominant
resolution term, due to non-compensation, unchanged.

\begin{figure}
 \centerline{\psfig{figure=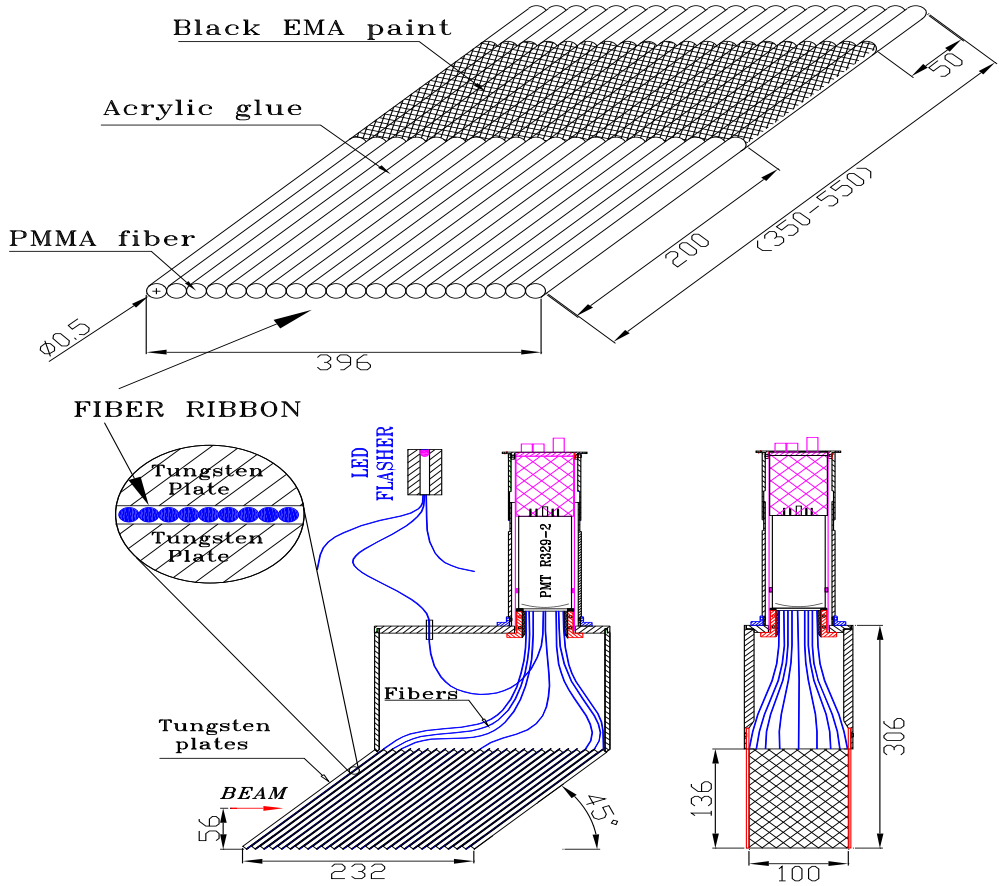,height=5in,width=6in}}
  \caption[mod]{ { \label{fg:mod}}
        Mechanical design of the production Tungsten Modules.Dimensions shown
are in mm.}
\end{figure}
\section{Module Construction}

For 10 cm wide modules with 5mm absorber plates, a convenient
longitudinal segmentation is 1 module per 2 nuclear interaction lengths
of absorber. The total fiber area matches that of a standard
$2^{\prime\prime}$ PMT\null.

For the prototype $W$ modules we obtained 2.5 mm thick cast plates from a
Russian manufacturer and bonded them in pairs. For the production
modules we obtained  machined plates of tungsten alloy with threaded
mounting holes from a US manufacturer\cite{Kulite}. The thickness uniformity
of our plates is $\pm$ 0.1mm. 

The fiber ribbons were wound on a mandrill
 and then impregnated with a low viscosity white
silicone rubber glue\cite{Ultraseal}. The glue covers the active region
of the fibers (200 mm) and protects the fiber surface in the
region of the fiber/absorber sandwich. The light guide section of the
remaining fibers is treated with an extramural absorber to suppress cladding modes in the fiber.
The fiber ends closest to the PMT are collected into an acrylic
compression fitting and impregnated with epoxy (Bicron BC-600)
. After the epoxy cured the fiber bundle was polished using a
diamond tipped cutting tool on a milling machine. 

The far end of the fibers were rough cut and left untreated. Our
optical simulations assume no reflection at this end.  

We removed 3 fibers at random from the ribbons in each module and
coupled them to a single external optical connector
for PMT gain monitoring. This allowed for stable optical connections of
all modules in the calorimeter stack to a single light flasher and
therefore reliable tracking of relative PMT gain.

As illustrated in Fig.~5, the fiber ribbons were trimmed to different
lengths depending on their positions along the module. Lengths were
adjusted to compensate for the difference in arrival time
between the front and back of the module. 
We kept the length of the acrylic fibers to a minimum because we were
concerned about additional light production in fibers outside the
absorber region- primarilly due to shower leakage at the top of the
calorimeter. 

We selected a 12-stage general purpose PMT (Hamamatsu R329-2)\cite{ham} and
mounted it with a 0.5mm air gap from the fiber bundles. PMT's were
 selected for $<6\%$ photocathode non-uniformity over the
39mm diameter area corresponding to the fiber bundle size. 
Linearity of the PMT/ voltage divider combination is also an important
criterion for this project since the calorimeters will be used to
measure up to 40 or so beam energy neutrons in collisions of gold ions.

\section{Precalibration}

All modules were tested for relative light yield using cosmic muons
incident along the beam axis and a standard PMT with calibrated response. Very
little variation ($<10\%$) was observed among the 24 tungsten modules we
eventually installed for this project. 

During testbeam operation we also installed trigger counters to select
a small contamination of muons which were present in the proton beam.
Muons traversing the full calorimeter were used to adjust relative
gains of the 4 PMT's in the prototype calorimeter. 

\section{Testbeam performance}

The calorimeters were mounted on a table with remote $x$-$y$ (transverse to
the beam) positioning in the CERN North area, downstream of experiment NA49. 
The beam size and position was defined to $\pm 1$ cm using a 1cm square
scintillation counter directly upstream of the table. 

The main purpose of the beam test was to study the response and
resolution of the calorimeter as a function of position. 

The $100$ and 160 GeV/c protons were selected
using a beam Cerenkov counter. The beam energy spread was typically
1\%. PMT current pulses were integrated and digitized using a standard
commercial ADC (LRS 2249w).

\begin{figure}
 \centerline{\psfig{figure=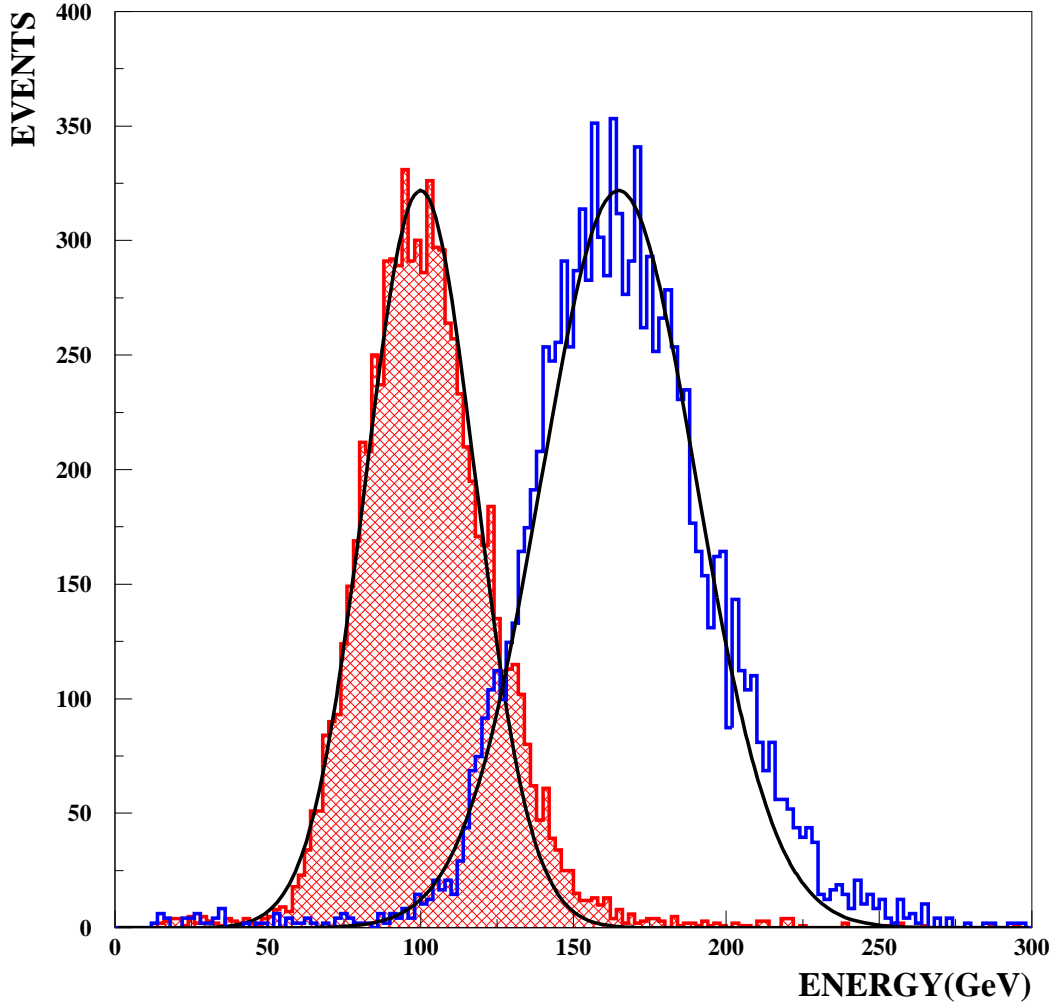,width=6in}}
  \caption[expresp]{ { \label{fg:expresp}}
        Tungsten ZDC response linshapes for 100 and 160 GeV incident protons. }
\end{figure}

\section{Results}

Figure~6 shows the measured lineshapes with 100 and 160 GeV incident
protons. Our energy scale is normalized using the 100 GeV point. 
Also shown in Fig.~6 are the expected distributions based on Geant
simulation. The distributions are well represented by a gaussian
resolution function and the response is linear over this limited energy
range.

\begin{figure}
 \centerline{\psfig{figure=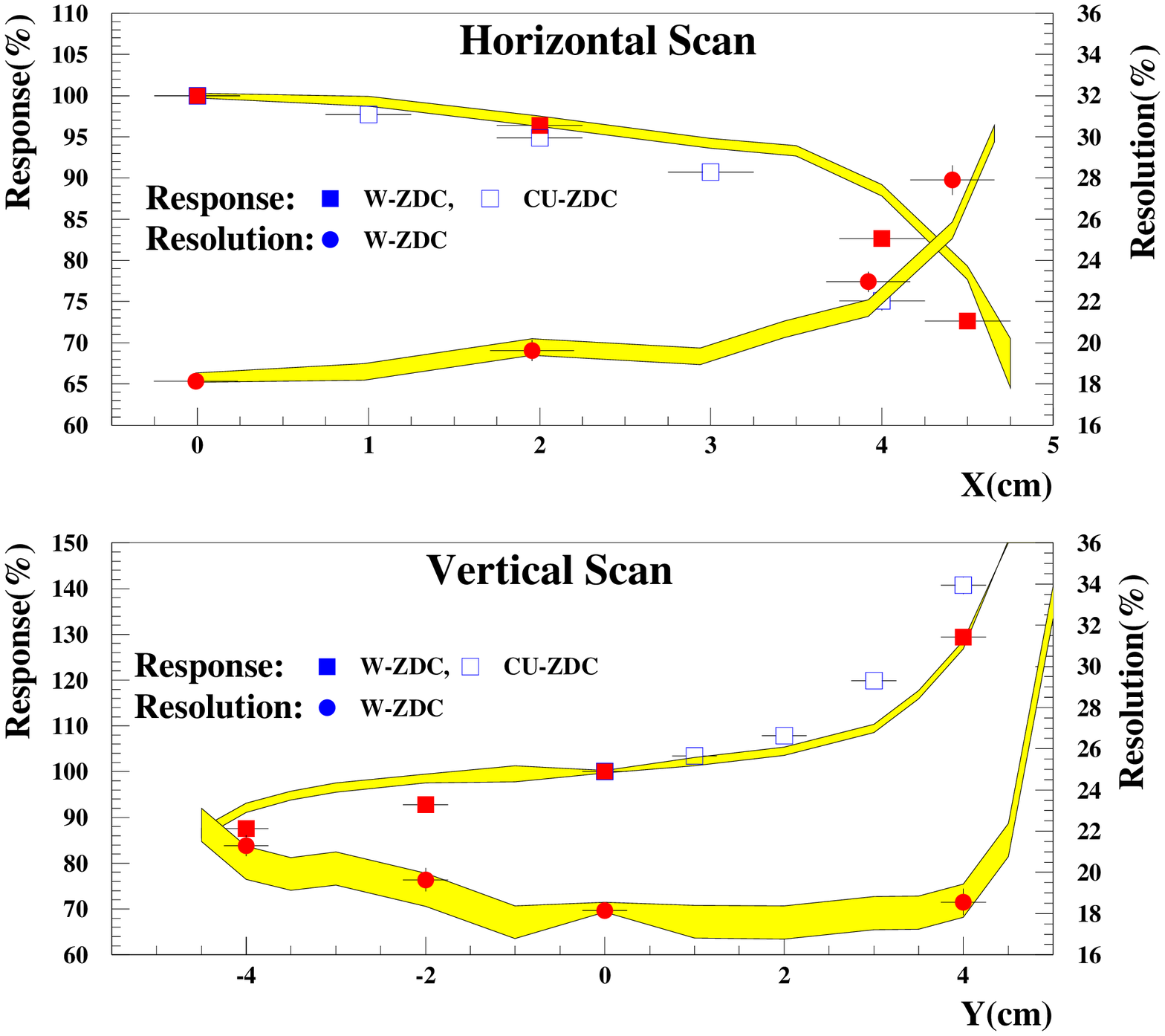,width=6in}}
  \caption[map]{ { \label{fg:map}}
        Response map of the calorimeters. }
\end{figure}

In Figure~7 we plot the prototype calorimeter response vs.\ impact position for
both $W$ and Cu module types. The position scans 
show essentially uniform
response to within 1cm of the calorimeter edge along the horizontal
direction -in good agreement with simulation. In the vertical scans
there is an abrupt increase in response near the upper edge of the
modules. Our simulation reproduces this edge effect. It can be traced
to shower leakage into the fibers above the absorber.

In order to improve module uniformity in the ``beam region'' we increased
the height of the module in our final production design from 10 cm to
13.6 cm. 

The tungsten calorimeter uniformity and energy resolution were
essentially unchanged when the energy deposited in the 4th module was   
neglected. Typically 1--2\% of the energy is seen in this module. The
energy resolution at 100 GeV changes from 17.6 to 19\% when it is removed.

In Figure~8 we plot the resolution (r.m.s./mean) for both prototypes
(using the full 8 $\Lambda_{I}$) at 100 and 160 GeV and compare them 
to 
simulation.
Part of the resolution of this calorimeter is
due to the unequal response to electrons and photons relative to hadrons.
This introduces a limiting resolution due to fluctuations in the shower
composition (ie $\pi^0$ vs charged $\pi$). Hence the non-zero intercept 
in Fig.~8.

\begin{figure}
 \centerline{\psfig{figure=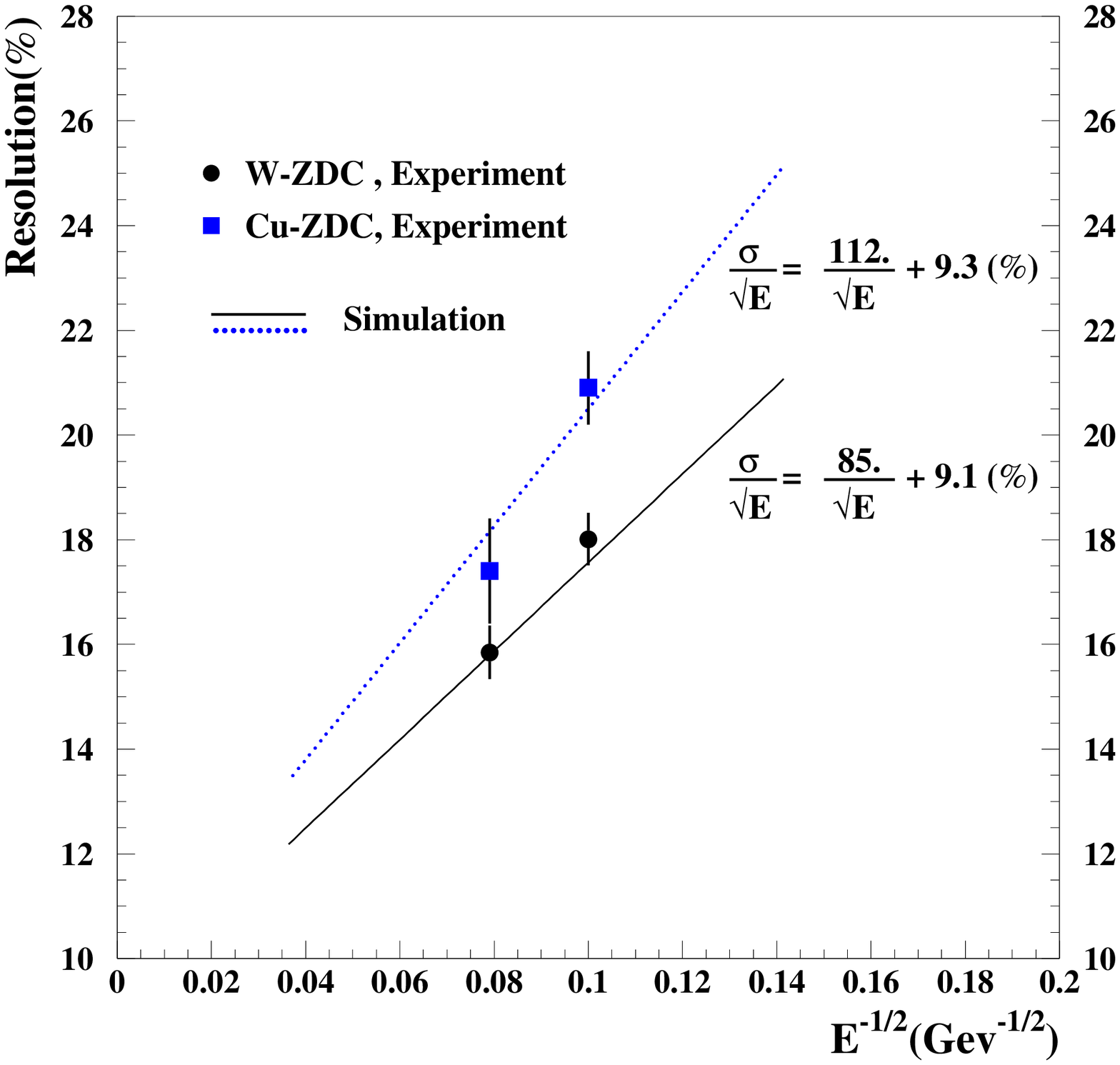,width=6in}}
  \caption[resol]{ { \label{fg:resol}}
        The ZDC energy resolution versus energy. }
\end{figure}

Our calorimeter is designed to measure beam energy neutrons incident at
the front face ($45^o$ to the fiber direction). A by-product of the
directional response of the calorimeter is that it is relatively
insensitive to background particles from ``beam halo'' and other sources.
In order to demonstrate this suppression, we inverted the calorimeter
in the testbeam ($135^o$ to fiber direction). 
The expected performance is shown in Fig.~9 together with the measured
response in the 2 configurations. The small discrepancy in the 
$135^o$ data could be accounted for by a 10\% reflection coefficient at
the open ends of the fiber.

\begin{figure}
 \centerline{\psfig{figure=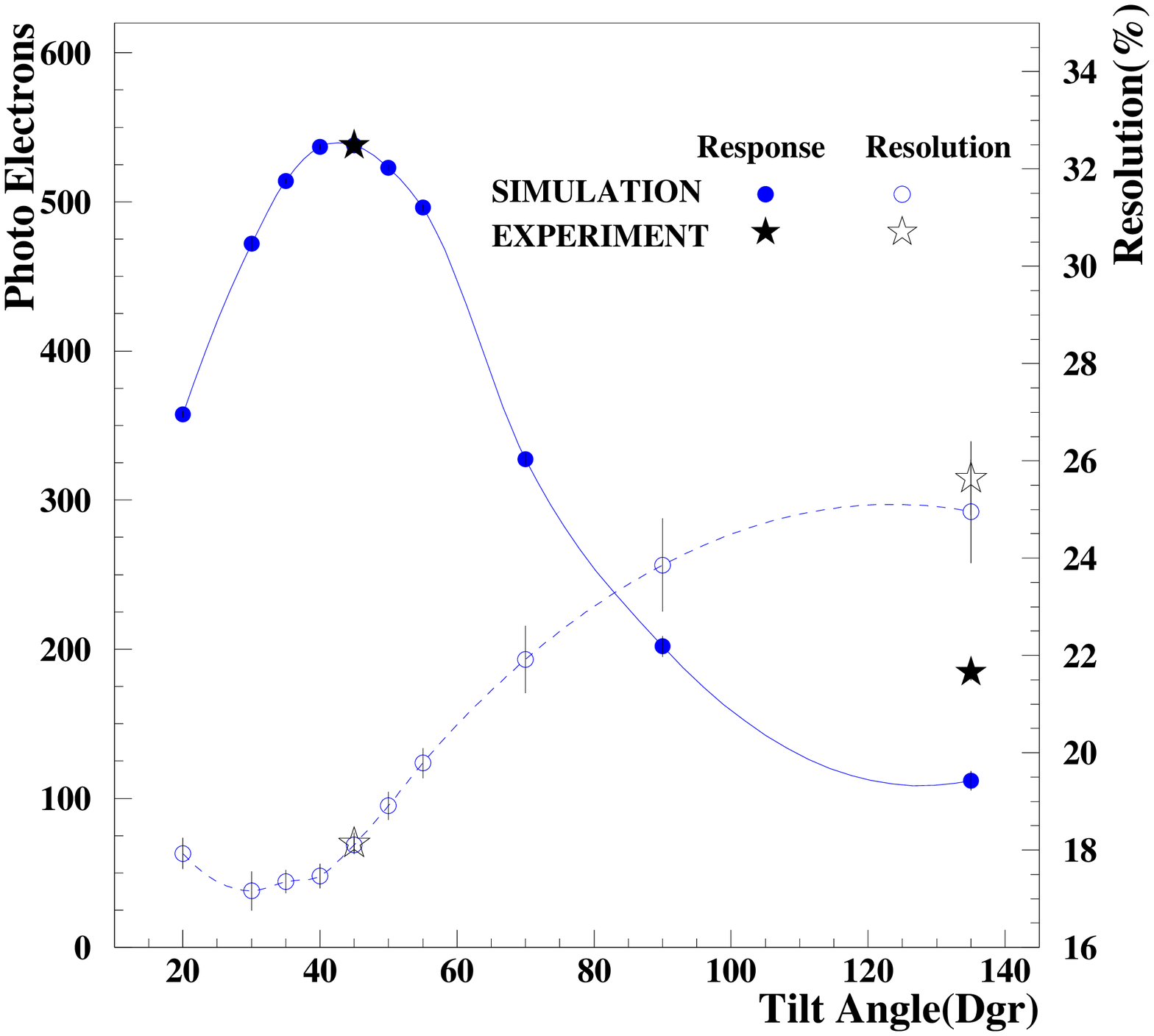,width=6in}}
  \caption[resol]{ { \label{fg:respang}}
        The ZDC response versus angle. }
\end{figure}

\subsection{Timing Resolution}

The ZDC signal formation is intrinsically fast. The main spread in
transit time of light emitted in the fibers results from propagation
along the fibers and depends on the vertical position of shower particles.
A design goal of the calorimeters is to achieve better than 200 psec
time-of-flight resolution. Time-of-flight measurement from the
interaction to each of the ZDC's is useful since it can be used to
determine the origin of beam interactions (which is proportional to the
time difference). 

In practice, the arrival time of the shower will be measured by
digitizing the time at which 1 or more PMT signals cross a threshold.
We corrected the measured time taking into account the signal risetime(2.5 nsec)
and the measured amplitude.

Fig.~10a shows a fit to the testbeam data from which we determine the
slew correction. In fig.~10b we plot the time difference between
signals from modules 1 \& 2. In figures~10c,d we plot the time
difference of the two module signals compared to the beam defining
scintillation counters. The resolution of the beam defining
scintillators is no better than that of the zdc but it is plausible ,
based on figure~10b, that the intrinsic ZDC resolution is $\sigma_t\leq 150 ps$
and it is certainly better than 200 psec. With a 150 psec
resolution the interaction point is measured to an error of 3 cms which
is perfectly adequate for our application.

\begin{figure}
 \centerline{\psfig{figure=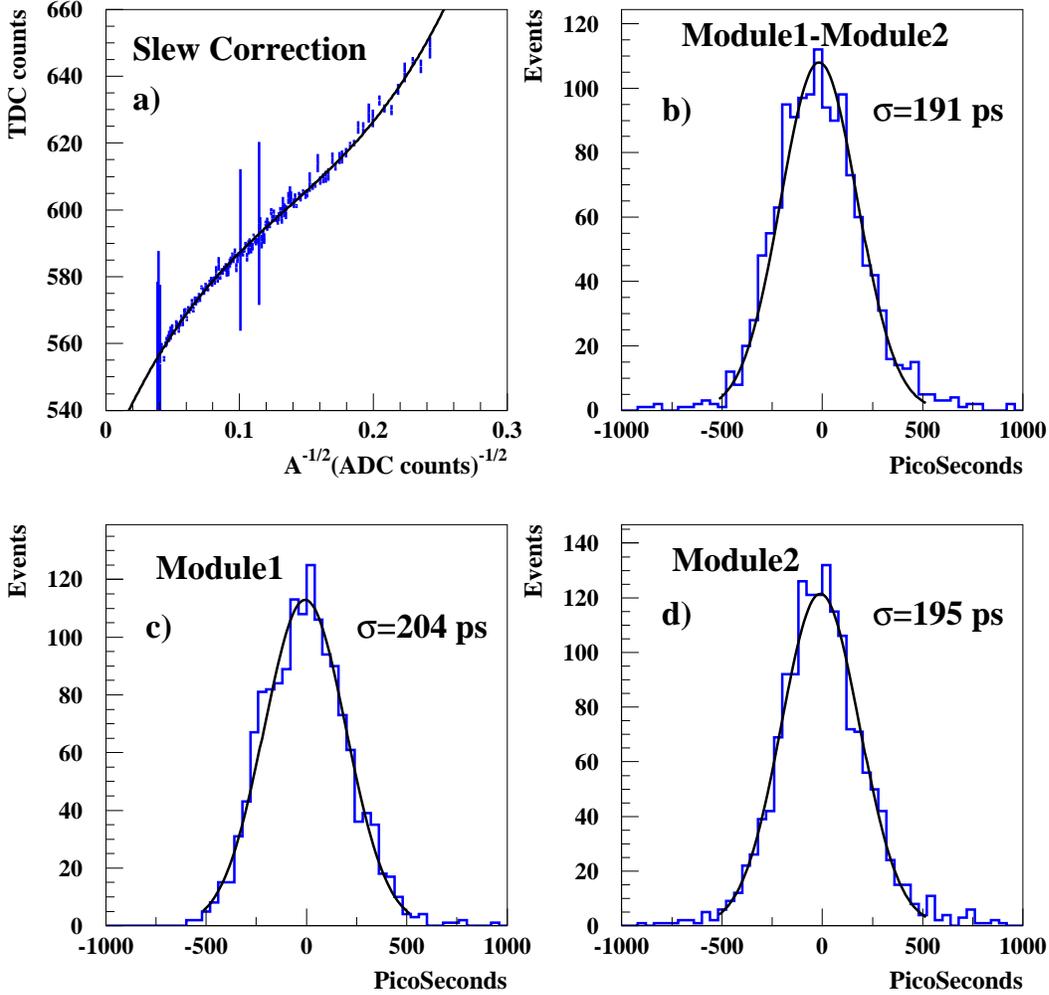,width=6in}}
  \caption[tres]{ { \label{fg:tres}}
        Time resolution. }
\end{figure}

\subsection{Radiation tolerance}

PMMA is not a particularly radiation tolerant plastic. It is known to
lose transparency more readily than Polystyrene, for example. Earlier
measurements on acrylic fibers \cite{brion} showed about a factor of 2
decrease in attenuation length per $10^4$ rad of gamma irradiation.
Doses at the ZDC location in RHIC have been estimated at 10krad/yr.
This is confirmed by dosimetry studies during RHIC commissioning.

\begin{figure}
 \centerline{\psfig{figure=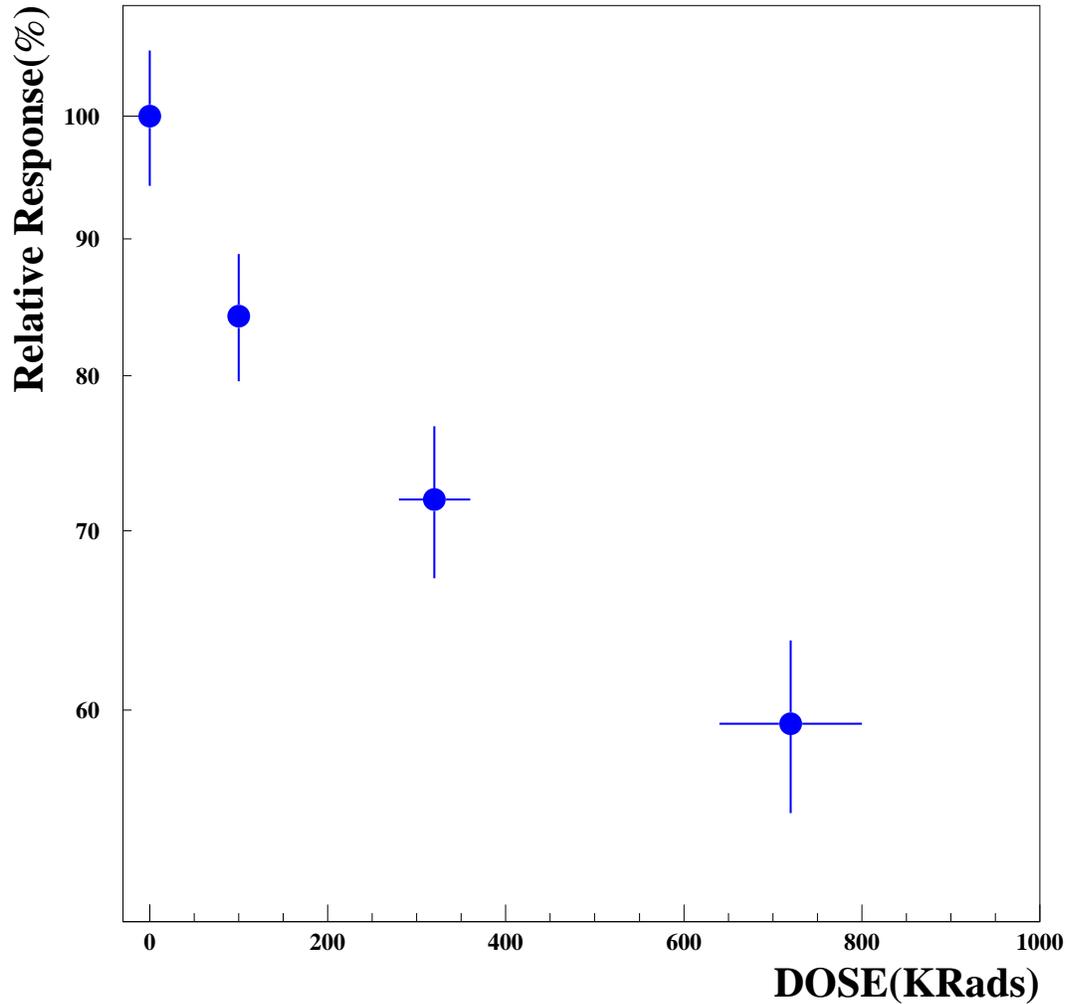,width=6in}}
  \caption[rad]{ { \label{fg:rad}}
        Shower response after after radiation exposure at a reactor. }
\end{figure}

We exposed one of our prototype modules to much higher integrated doses
at a reactor\cite{tamu}. Approximately 2/3 of the total dose was due to 
gamma rays and the remainder was due to neutrons
 The module light output was measured using cosmic ray muons before and
after 3 successive exposures up to a maximum of 700 krad (7000Gy). The
results are plotted in fig.~13, from which we conclude that the useful
lifetime of the calorimeters will be $> 500$ krad.  

\section{Production Design Choices}
	
The Copper and Tungsten modules both had adequate performance for our 
application. The tungsten module yields 1-2\% better energy resolution at
100 GeV and slightly better flatness of response over the calorimeter 
face. We chose to proceed with the Tungsten module design primarily
because of the $2\times\Lambda_{I}$ modularity and other aspects of
the  mechanical design. Also the modules are more compact, which is an 
advantage given our limited space.
The module height was increased in the production design to reduce shower
leakage into the fiber bundles. Also the calorimeter depth was reduced from 
8 to $6\times\lambda_{I}$ since the shorter calorimeter gives essentially identical 
performance.

\begin{figure}                                                          
 \centerline{\psfig{figure=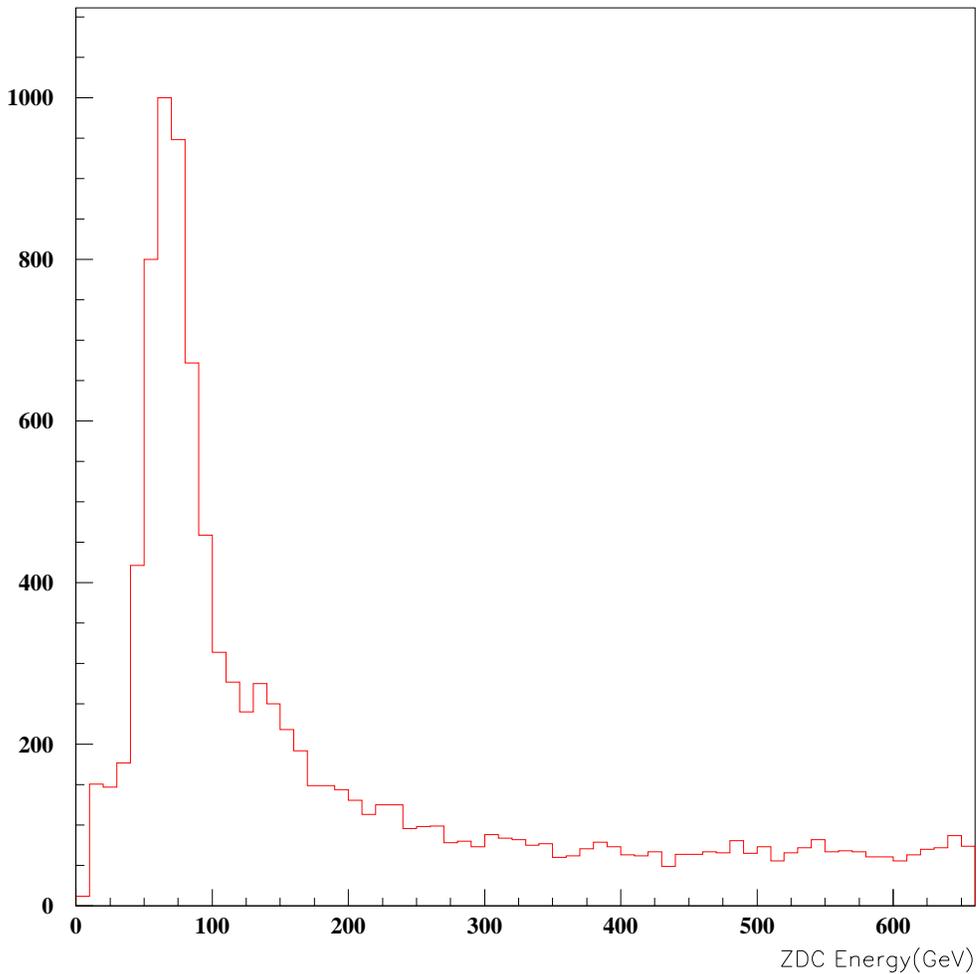,width=6in}}                 
  \caption[zdcdist]{ { \label{fg:zdcdist}}                                      
       Online Zero Degree Calorimeter Energy Distribution obtained during
RHIC colliding beam operation with beam energies of 65 GeV/nucleon. }  
\end{figure}                                                            
 
\section{Discussion}

Studies of the Cerenkov/fiber sampling technique with electrons \cite{Lundin}
 and have been reported elsewhere. One device
 \cite{Musso} was built and operated in a fixed target experiment. Here
we report first measurements with a $45^o$ design hadron calorimeter
and the first application of this method at a collider.

	During the first colliding beam operation at RHIC, with Gold Ions 
accelerated to and stored at 65GeV/nucleon beam energy, the ZDC's have been 
used for beam tuning and as a trigger by the RHIC experiments. Figure ~12 shows
an online Energy distribution from one of the calorimeters in the PHENIX
experiment. The single neutron peak, which is seen clearly in this distribution,
has been used to confirm the energy calibration of the calorimeters.
\section{Acknowledgements}

We wish to thank Phillipe Gorodetzky for many useful discussions
throughout this project. We also thank Alice Mignerey and Bill Christie
for assistance at  various stages. The kind assistance of the NA49
collaboration which allowed us to share their beamline and counting house
is
also acknowledged.
This work has been partially supported under DOE grant \#\underbar 
DE-AC02-98CH10886 and DE-FG03-93ER40773
{\hskip1.2in}.

\end{document}